\documentclass[fleqn,usenatbib]{mnras}

\usepackage{newtxtext,newtxmath}
\usepackage[T1]{fontenc}
\usepackage{ae,aecompl}
\usepackage{url}
\usepackage{graphicx}
\usepackage{xspace}
\usepackage{amsfonts}
\usepackage{pifont}
\usepackage{textcomp}
\usepackage{float}
\usepackage{subfig}
\usepackage{multirow}
\usepackage{adjustbox}

\title[Spectra of Ultra Fast Rotators]{The Puzzling Story of Flare Inactive Ultra Fast Rotating M dwarfs. II. Searching for radial velocity variations}

\author[G. Ramsay et al.]{
Gavin Ramsay$^{1}$\thanks{E-mail: gavin.ramsay@armagh.ac.uk}, Pasi Hakala$^{2}$, J. Gerry Doyle$^{1}$, Lauren Doyle$^{3,4}$, Stefano Bagnulo$^{1}$
\\
$^{1}$Armagh Observatory and Planetarium, College Hill, Armagh, BT61 9DG, UK\\
$^{2}$Finnish Centre for Astronomy with ESO (FINCA), Quantum, University of Turku, FI-20014, Finland\\
$^{3}$Centre for Exoplanets and Habitability, University of Warwick, Coventry, CV4 7AL, UK \\
$^{4}$Department of Physics, University of Warwick, Coventry, CV4 7AL, UK\\
}

\date{Accepted 2022 January 18. Received 2022 January 17; in original form 2021 October 11}

\begin{document}
\label{firstpage}
\pagerange{\pageref{firstpage}--\pageref{lastpage}}

\outer\def\gtae {$\buildrel {\lower3pt\hbox{$>$}} \over 
{\lower2pt\hbox{$\sim$}} $}
\outer\def\ltae {$\buildrel {\lower3pt\hbox{$<$}} \over 
{\lower2pt\hbox{$\sim$}} $}
\newcommand{\Msun}{$M_{\odot}$}
\newcommand{\lsun}{$L_{\odot}$}
\newcommand{\Rsun}{$R_{\odot}$}
\newcommand{\solar}{${\odot}$}
\newcommand{\kep}{\sl Kepler}
\newcommand{\ktwo}{\sl K2}
\newcommand{\tess}{\sl TESS}
\newcommand{\swift}{\it Swift}
\newcommand{\Porb}{P_{\rm orb}}
\newcommand{\nuorb}{\nu_{\rm orb}}
\newcommand{\eplus}{\epsilon_+}
\newcommand{\eminus}{\epsilon_-}
\newcommand{\cd}{{\rm\ c\ d^{-1}}}
\newcommand{\MdotL}{\dot M_{\rm L1}}
\newcommand{\Mdot}{$\dot M$}
\newcommand{\Mdsolar}{\dot{M_{\odot}} yr$^{-1}$}
\newcommand{\Ldisk}{L_{\rm disk}}
\newcommand{\src}{KIC 9202990}
\newcommand{\ergscm} {erg s$^{-1}$ cm$^{-2}$}
\newcommand{\rchi}{$\chi^{2}_{\nu}$}
\newcommand{\chisq}{$\chi^{2}$}
\newcommand{\pcmsq} {cm$^{-2}$}

\maketitle

\begin{abstract}
Observations made using {\tess} revealed a sample of low mass stars
which show a periodic modulation on a period $<0.2$~d. Surprisingly
many of these Ultra Fast Rotating (UFR) stars showed no evidence of
flare activity which would be expected from such rapidly rotating
stars. We present results from a spectroscopic survey of UFRs using
the Nordic Optical Telescope to search for radial velocity variations
which could reveal evidence for binarity. Our sample of 29 sources
have a photometric period between 0.1-0.2d, cover spectral classes of
M0-4V, and show no evidence for flares.  We detect only one source
with clear radial velocity shifts, with another two having Gaia
  RUWE values which suggests they are binaries. Further observations
reveal the former star possibly contains a brown dwarf companion with
a mass of M$_2$>58 M$_{\rm Jup}$ and probability P(M$_2$<90 M$_{\rm
  Jup}$) = 50\%. There is no evidence for the companion in our
spectra, strengthening the case for a brown dwarf companion. We also
examine the folded {\tess} light curves of all our targets, finding at
least two are eclipsing binaries and one which has been contaminated
by a spatially nearby $\delta$ Sct star. We estimate that around 1/4
of our targets may have been contaminated by short period variable
stars. However, the majority of our targets are consistent with being
single, low mass stars whose variability is due to starspots. We
outline the possible reasons why they are not flare active despite
being such rapid rotators.

\end{abstract}


\section{Introduction}

It has long been known that the rotation period of stars increases as
they age \citep[e.g.][]{Skumanich1972}. However, until recently
determining the precise rotation period of stars was a laborious and
time consuming process. This is because stars were generally observed
on a star-by-star basis and had data gaps introduced by diurnal and
poor weather effects, which is especially true for stars with
rotational periods longer than $\sim$4--5 hours. Matters changed with
the launch of {\kep} \citep{Borucki2010} which provided the data to
measure the rotation period of tens of thousands of stars along the
main sequence \citep[e.g.][]{McQuillan2014}. After the initial four
year mission, {\kep} made a series of observations of fields along the
ecliptic plane each lasting several months, with the mission being
re-named {\sl K2}. Studies of open clusters of different ages were
able to determine in more detail how the stellar rotation period
varies as a function of age and mass
\citep[e.g.][]{Rebull2016}. However, binarity can also effect the
rotation rate of stars. For example, in a study of stars in the Open
Cluster Blanco 1, \citet{Gillen2020} showed that mid-F to mid-K stars
which were in binaries have faster rotation rates than single stars of
the same type. This suggests the companion reduces angular momentum
loss even at ages of $\sim$100 Myr.

As stellar activity is related to a stars rotation period
\citep{HartmannNoyes1987,Yang2017}, stars become less magnetically
active as they age \citep[see][ and references
  within]{Davenport2019}. Stellar activity can manifest itself in
different ways including: starspots, narrow optical line emission,
X-ray emission and flare activity. Although flares have been seen on
stars with earlier spectral-types, they appear more common on low
mass, fully convective (later than $\sim$M3/4) stars in particular
\cite[e.g.][]{Pettersen1989}.

Optical flares have been studied on low mass stars using {\kep}
\citep[e.g.][]{Ramsay2013,Hawley2014} and {\ktwo}
\citep[e.g.][]{Ramsay2015,Gizis2017,Doyle2018}. The launch of {\tess}
in April 2018 opened up a window on nearly the whole sky and allowed
at least month long photometric observations with 2-min cadence for
tens of thousands of stars \citep{Ricker2015}.  In a study utilising
2-min cadence lightcurves from {\tess}, \citet{Doyle2019} conducted an
analysis of stellar flares on 149 M dwarfs. During our
study, a small group of low mass Ultra Fast Rotating stars (UFRs) were
identified which have rotation periods $<$0.3 d and show low
levels of flaring activity. We did not find evidence that the lack of
activity is related to stellar age or rotational
velocities. Similarly, \citet{Gunther2020} used data taken from the
first two months of the {\tess} mission and found there was a
`tentative' decrease in the flare rate for stars with $P<$
0.3~d. Given that fast rotating stars should display high levels
of activity, why do these rapidly rotating stars show little to no
flaring activity?

To address this question further, we made a systematic search for UFRs
using all the southern ecliptic 2-min cadence data in
\citet{Ramsay2020}. Out of 9887 stars brighter than $T$=14 mag and
close to the main sequence, 609 were found to be low mass stars with a
period $<$1~d. Of these, only 288 showed at least one flare. For stars
with periods $>$0.4~d, 51\% of stars are flare active, whilst for
stars with periods $<$0.2~d the fraction is 11\%. Overall, these
findings from \citet{Ramsay2020} strengthened the initial findings of
\citet{Doyle2019} and \citet{Gunther2020}.

In \citet{Doyle2019} and \citet{Ramsay2020} we suggested several
reasons why the majority of low mass stars with rotation periods
$<$0.2~d do not appear to show optical flares:

\begin{enumerate}

\item They {\sl do} show low-energy flares, perhaps at bluer
  wavelengths that would not be detected using {\tess} (which has a
  response between 6000--10000~\AA). For instance \citet{Namekata2020}
  show multi-band observations of AD Leo in which one flare was seen
  in $g$ and $R$ bands but not $i$. High-cadence photometry of low
  mass stars with a $P <$0.2~d, especially in the $U$ band, could reveal
  these `missing' flares.

\item The binary system GJ 65, contains two variable stars (UV Cet and
  BL Cet) which possess dramatically different magnetic field strength
  and configurations along with varying degrees of activity at
  different energies \citep{Kochukhov2017}. \citet{Shulyak2017}
    find evidence that low mass stars with simple dipole fields can
    have the strongest magnetic fields, whilst those stars with
    multipole fields cannot generate fields stronger than $\sim$4
    kG. This suggests the magnetic field configuration of the stars
    plays an important role in their magnetic activity, perhaps more
    so than their rotation period or age. Therefore, could the
    magnetic field configuration be the cause of those UFR showing no
    or few optical flares?
  
\item The {\tess} pixels are 21$^{"}$ square implying that light
  from spatially nearby stars may dilute the light from the target. If
  a variable star were nearby then this could contaminate the light
  curve of the target making the target variable on an unrelated
  period and also dilute or mask any flares from the target. Similarly
  other binary stars such as short period contact binaries or
  cataclysmic variables have light curves which resemble those
  expected from isolated low mass stars with starspots.

\item Binary stars with orbital periods \ltae4 d are likely to be
  synchronised \citep{Lurie2017,Fleming2019} with stars with the
  shortest periods likely to be non-spherical due to the tidal
  force. It is, therefore, possible that the period we detect in the
  {\tess} data could be a signature of an orbital period rather than
  the rotation period of a single star. However, it is not clear why
  the magnetic activity of both binary components would be suppressed,
  given that they both co-rotate with the same short period.
  
\end{enumerate}

To explore these issues further, in a companion paper
\citep[][hereafter referred to as Paper I]{Doyle2022}, we used the
VLT/FORS2 instrument to make spectropolarimetric observations of ten
UFRs and found that five had a line of sight magnetic field
$\sim$1--2~kG. However, with only half of our sample having a
detectable line of sight magnetic field, and four of those being
the more active stars in the sample, it would appear the magnetic
field strength may not be the answer to the lack of flaring activity
in UFRs. We note, however, that FORS2 low-resolution
  spectropolarimetry is only sensitive to the component of the
  magnetic field along the line of sight, averaged over the stellar
  disk. This quantity may be very small or null, even in the presence
  of a relatively strong surface field. The lack of detection with
  FORS2 cannot be used to rule out the present of a magnetic field
  with a complex morphology. Stronger conclusions could be reached
  with high S/N, high-resolution spectropolarimetry, exploiting the
  fact that regions of the stellar disk characterised by different
  field strength may have different radial velocities, due to stellar
  rotation, and may be responsible each of them for local Stokes $V$
  profiles centred at different wavelengths \citep[this is the well
    known "cross-over" phenomenon already discovered
    by][]{Babcock51}. The average Stokes profiles would still have a
  null zero-order moment about the line centre (to be interpreted as a
  zero mean longitudinal field), but the presence of a magnetic field
  could be revealed by ripples on the Stokes $V$ profile that would
  pass undetected at lower spectral resolution.

In this paper, we search for evidence for binarity in a sample
of UFRs.  To do this we use the Nordic Optical Telescope (NOT) to
obtain spectra of a sample of UFRs made over three days to search for
radial velocity variations.

\section{Selection of Targets using {\tess} data}

In \citet{Ramsay2020} we reported the results of a search of UFRs in
the southern ecliptic hemisphere ({\tess} Cycle 1) using {\tess} 2 min cadence
data. In this paper we report on a similar study using data from the
northern ecliptic hemisphere ({\tess} Cycle 2). 

\subsection{Determining Periods and searching for Flares}

In summary, we downloaded the calibrated lightcurves of our targets
from the MAST data
archive\footnote{\url{https://archive.stsci.edu/tess/}}. We used the
flux values for {\tt PDCSAP\_FLUX}, which are the Simple Aperture
Photometry values, {\tt SAP\_FLUX}, after correction for systematic
trends. We removed photometric points which did not have {\tt
  QUALITY=0} flag. To determine the rotation period of the stars, we
used the generalised Lomb Scargle
\citep[LS,][]{Zechmeister2009,Press1992} and Analysis of Variance
\citep[AoV,][]{Schwarzenberg1996} periodograms to identify the most
prominent period in each of the stars light curves from each
sector. The results from the LS and AoV periodograms were consistent
although the significance of the main period could vary between
different sectors.
  
To search for flares in the light curves, we removed the signature of
the rotational modulation using a routine in the {\tt lightkurve}
package \citep{lightkurve2018}. We then searched these flattened light
curves for flares using the {\tt
  Altaipony}\footnote{\url{https://altaipony.readthedocs.io/en/latest}}
suite of software which is an update of the {\tt Appaloosa}
\citep{Davenport2016} software package.
  
In selecting targets to be observed using the NOT we had four main
criteria: their visibility from La Palma at the time of the
observations; they showed a clear periodic modulation in their {\tess}
2 min cadence light curve; were $i<$14 mag \citep[we used the
  Pan-STARRS DR2 catalogue][]{Chambers2016} and had a position in the
Gaia HRD \citep{Gaia2021} which was close to the main sequence, so we
did not target stars which were either very young or likely binary
stars.

In Table \ref{targets} we show the targets, the number of sectors
where the source was observed in Cycle 2, and include the period which
we derived from {\tess} 2 min data. All targets have the most
prominent peak in their LS and AoV periodogram $<$0.2 d and none show
optical flares.

\begin{table*}
\caption{Details of the targets in our sample. We show their {\tess}
  Input Catalogue ID (TIC, \citet{Stassun2019}); RA and DEC taken from
  the TIC; the $i$ mag taken from Pan-STARRS DR2 \citep{Chambers2016};
  the number of Sectors in which the star was observed in 2 min
  cadence mode during Cycle 2; the period; the semi-amplitude
  expression as fraction (both determined from {\tess} observations);
  the $(BP-RP)$ colour and $M_{G}$ absolute $G$ mag \citep{Gaia2021};
  the effective temperature taken from the TIC and expressed to 3
  significant figures (the quoted uncertainty is 157 K) and the
  Spectral Type determined using the $G-G_{RP}$ colour
  \citep{Kiman2019}. In the notes column, EB refers to eclipsing
  binary and RV mod indicates it shows a radial velocity modulation,
  whilst U and X indicate the source was detected in the Galex all-sky
  survey \citep{Bianchi2017} and the Rosat All-sky Survey Faint Source
  Catalogue \citep{Voges2000} respectively.}
\begin{tabular}{rllrrrccrrrl}
\hline 
  \multicolumn{1}{c}{TIC} &
  \multicolumn{1}{c}{RA} &
  \multicolumn{1}{c}{DEC} &
  \multicolumn{1}{c}{$i$} &
  \multicolumn{1}{c}{\#Sectors} & 
  \multicolumn{1}{c}{Period} &
  \multicolumn{1}{c}{Amplitude} &
  \multicolumn{1}{c}{$(BP-RP)$} &
  \multicolumn{1}{c}{$M_{G}$} &
  \multicolumn{1}{c}{$T_{eff}$} &
  \multicolumn{1}{c}{SpT}&
  \multicolumn{1}{l}{Notes} 
 \\
  \multicolumn{1}{c}{} &
  \multicolumn{1}{c}{(J2000)} &
  \multicolumn{1}{c}{(J2000)} &
  \multicolumn{1}{c}{(mag)} &
  \multicolumn{1}{c}{} &
  \multicolumn{1}{c}{(d)} &
  \multicolumn{1}{c}{fraction} &
  \multicolumn{1}{c}{} &
  \multicolumn{1}{c}{} &
  \multicolumn{1}{c}{(K)} &
  \multicolumn{1}{c}{} &
  \multicolumn{1}{c}{}
  \\
\hline
  452912864 & 00:33:14.8 & +55:55:21.5 & 13.7 & 1 & 0.054 & 0.0130 & 2.19 &  9.13 & 3630& 1.6 & \\
  421117621 & 00:46:11.8 & +63:03:20.6 & 13.5 & 3 & 0.159 & 0.0046 & 2.27 &  9.49 & 3570& 1.9 & \\
  285039638 & 00:48:43.8 & +61:16:42.6 & 13.0 & 2 & 0.168 & 0.0157 & 1.90 &  8.02 & 3860& 0.7 & \\
  351876189 & 00:54:48.6 & +66:07:30.0 & 12.7 & 3 & 0.175 & 0.0039 & 2.00 &  8.55 & 3770& 1.0 & \\
  288500817 & 02:34:14.9 & +45:42:38.8 & 13.1 & 1 & 0.161 & 0.0091 & 2.70 & 10.56 & 3330& 3.1 & U \\
  418207289 & 03:19:44.2 & +72:51:57.2 & 13.8 & 1 & 0.147 & 0.0099 & 2.46 & 10.18 & 3460& 2.5 & \\
  418208790 & 03:19:54.3 & +74:49:12.9 & 13.1 & 1 & 0.180 & 0.0039 & 2.00 &  8.93 & 3770& 1.0 & \\
  354790015 & 03:31:45.5 & +49:42:37.4 & 12.7 & 1 & 0.063 & 0.0052 & 1.90 &  8.21 & 3860& 0.6 & \\
  256738604 & 03:51:26.5 & +82:38:44.6 & 13.2 & 1 & 0.148 & 0.0035 & 2.60 & 10.24 & 
  3380& 2.9 & U \\
  187254179 & 04:49:42.5 & +39:35:03.5 & 12.6 & 1 & 0.170 & 0.0093 & 1.96 &  8.21 & 3810& 0.9 & \\
  281571049 & 04:49:55.7 & +71:09:47.1 & 11.9 & 1 & 0.174 & 0.0103 & 2.70 & 10.43 & 3330 & 3.1 & \\
  327871640 & 05:08:38.8 & +49:56:33.8 & 13.3 & 1 & 0.169 & 0.0049  & 2.76 & 11.02 & 3300& 3.2 & \\
  310162555 & 05:30:00.5 & +51:08:49.1 & 13.9 & 1 & 0.161 & 0.0118 & 2.81 & 11.13 & 3270 & 3.3 & \\
  116609201 & 05:42:22.6 & +34:52:44.0 & 13.5 & 1 & 0.161 & 0.0505 & 1.94 &  8.08 & 3830 & 0.9 & $\delta$ Sct like\\
  155657579 & 13:45:54.3 & +79:23:15.0 & 13.8 & 2 & 0.184 & 0.0077 & 2.74 & 10.58 & 3300 & 3.2 & EB\\
  85407625  & 17:10:11.0 & +41:39:34.2 & 11.7 & 2 & 0.176 & 0.0329 & 2.45 &  9.11 & 3460& 2.4 & RV mod, X, U\\
  329248235 & 17:35:32.4 & +54:27:36.4 & 13.2 & 9 & 0.128 & 0.0022 & 2.57 &  9.90 & 3390& 2.7 & \\
  258922572 & 19:14:36.1 & +69:28:51.6 & 12.9 & 12 & 0.183 & 0.0020 & 2.10 &  8.76 & 3690& 1.3 & U \\
  282773740 & 20:06:42.2 & +19:21:42.5 & 11.2 & 1 & 0.188 & 0.0014 & 2.03 &  8.62 & 3750& 1.1 & \\
  387330194 & 20:39:34.5 & +68:22:12.7 & 12.6 & 6 & 0.192 & 0.0030 & 2.05 &  8.31 & 3730& 1.2 & U \\
  136513953 & 21:09:50.1 & +42:57:20.2 & 13.8 & 2 & 0.137 & 0.0200 & 2.66 & 10.66 & 3350 & 3.0 & \\
  137188834 & 21:12:30.3 & +42:55:35.1 & 13.3 & 2 & 0.154 & 0.0068 & 1.92 &  7.89 & 3840& 0.7 &  \\
  429916899 & 21:33:12.8 & +55:01:23.0 & 12.6 & 1 & 0.142 & 0.0159 & 1.85 &  7.86 & 3910 & 0.5 & \\
  394885751 & 21:39:17.5 & +49:24:47.2 & 13.0 & 2 & 0.164 & 0.0051 & 1.95 &  8.17 & 3820 & 2.0 & \\
  419666455 & 21:40:18.0 & +41:08:03.5 & 13.7 & 2 & 0.197 & 0.0033 & 2.39 &  9.77 & 3500 & 2.2 & EB \\
  346130527 & 21:49:56.0 & +47:25:48.7 & 12.8 & 2 & 0.187 & 0.0041 & 1.99 &  8.27 & 3780 & 1.0 & \\
  66635046 & 22:46:13.7 & +48:15:12.6  & 13.3 & 2 & 0.191 & 0.0073 & 2.60 & 10.65 & 3380 & 2.8 & \\
  279606560 & 23:06:00.5 & +71:42:31.4 & 13.5 & 1 & 0.178 & 0.0070 & 2.31 &  8.85 & 3540 & 2.9 & \\
  251922596 & 23:28:19.2 & +58:12:19.8 & 13.1 & 2 & 0.159 & 0.0113 & 2.08 &  8.39 & 3700 & 1.3 &  EB?\\
\hline\end{tabular}
\label{targets}
\end{table*}

\begin{figure}
  \begin{center}
  \includegraphics[width=0.45\textwidth]{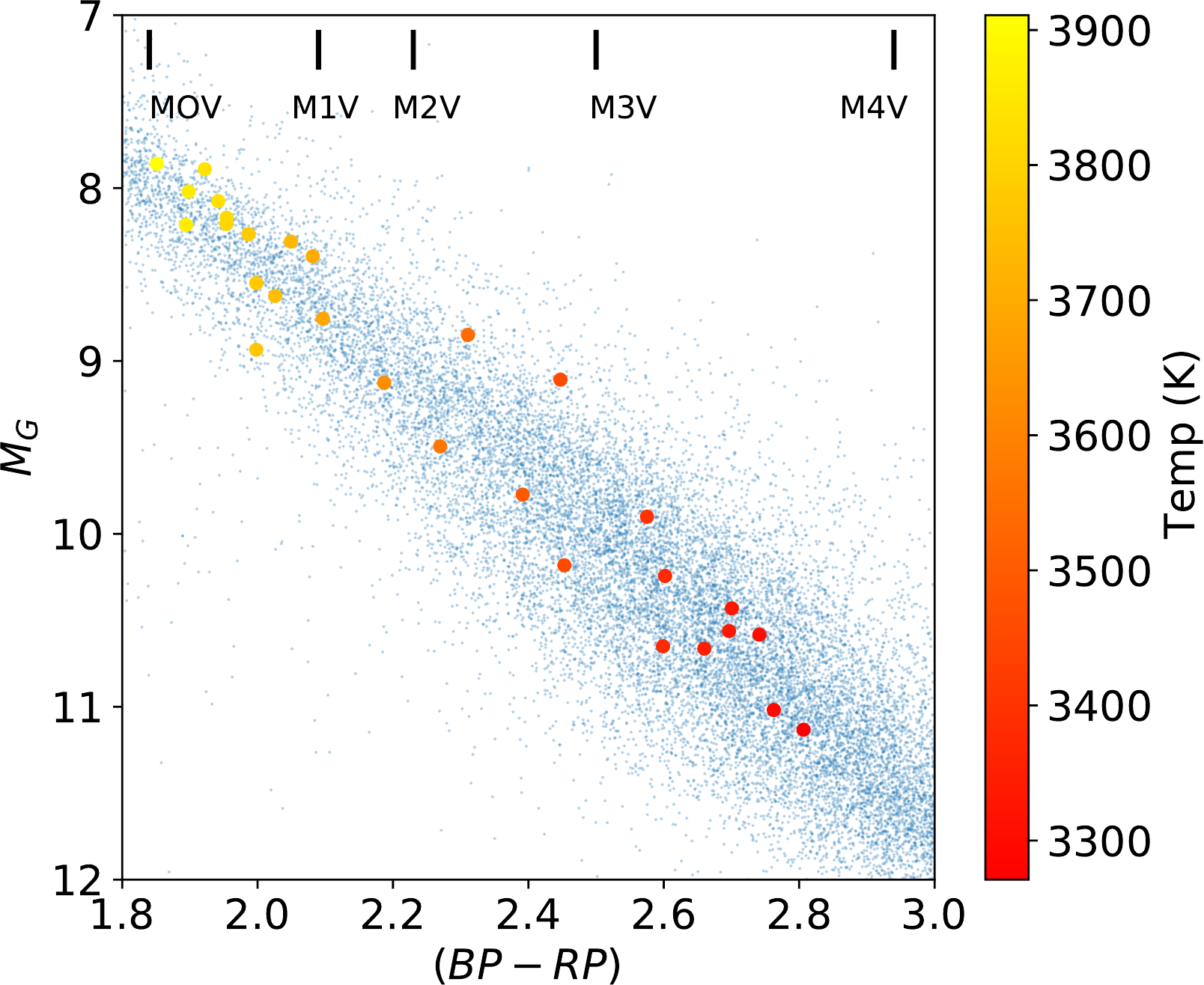}    
\vspace{2mm}
  \caption{The Gaia HRD ($B-R$, $M_{G}$) where the small dots come
    from stars within 50 pc of the Sun and our targets are shown as
    larger dots \citep{Gaia2021} and their colour reflects their
    temperature derived from the TIC \citep{Stassun2019}. The colour
    of spectral sub-types has used the work of
    \citet{PecautMamajek2013}.}.
    \label{gaiahrd}
    \end{center}
\end{figure}

\subsection{Targets}

In Figure \ref{gaiahrd} we show the location of our targets in the
Gaia HRD, i.e. in the $(BP-RP),M_{G}$ plane, which implies they have
spectral types in the range
$\sim$M0-M4V\footnote{\url{https://www.pas.rochester.edu/~emamajek/EEM_dwarf_UBVIJHK_colors_Teff.txt}}
\citep{PecautMamajek2013}. The spectral type determinations are
  broadly consistent with those determined using the Gaia $G-G_{RP}$
colour relationship with spectral type \citep{Kiman2019} which we show
in Table \ref{targets}. All targets are located very close to the main
locus of the main sequence, with only three being slightly offset. We
also searched Galex all-sky survey in the UV \citep{Bianchi2017} and
the Rosat all-sky Faint Catalogue Survey in soft X-rays
\citep{Voges2000} and find five UV matches (indicated as a 'UV' in
Table \ref{targets} and one X-ray match (TIC 85407625).

\subsection{Folded light curves}

\begin{figure*}
  \begin{center}
  \includegraphics[width=0.8\textwidth]{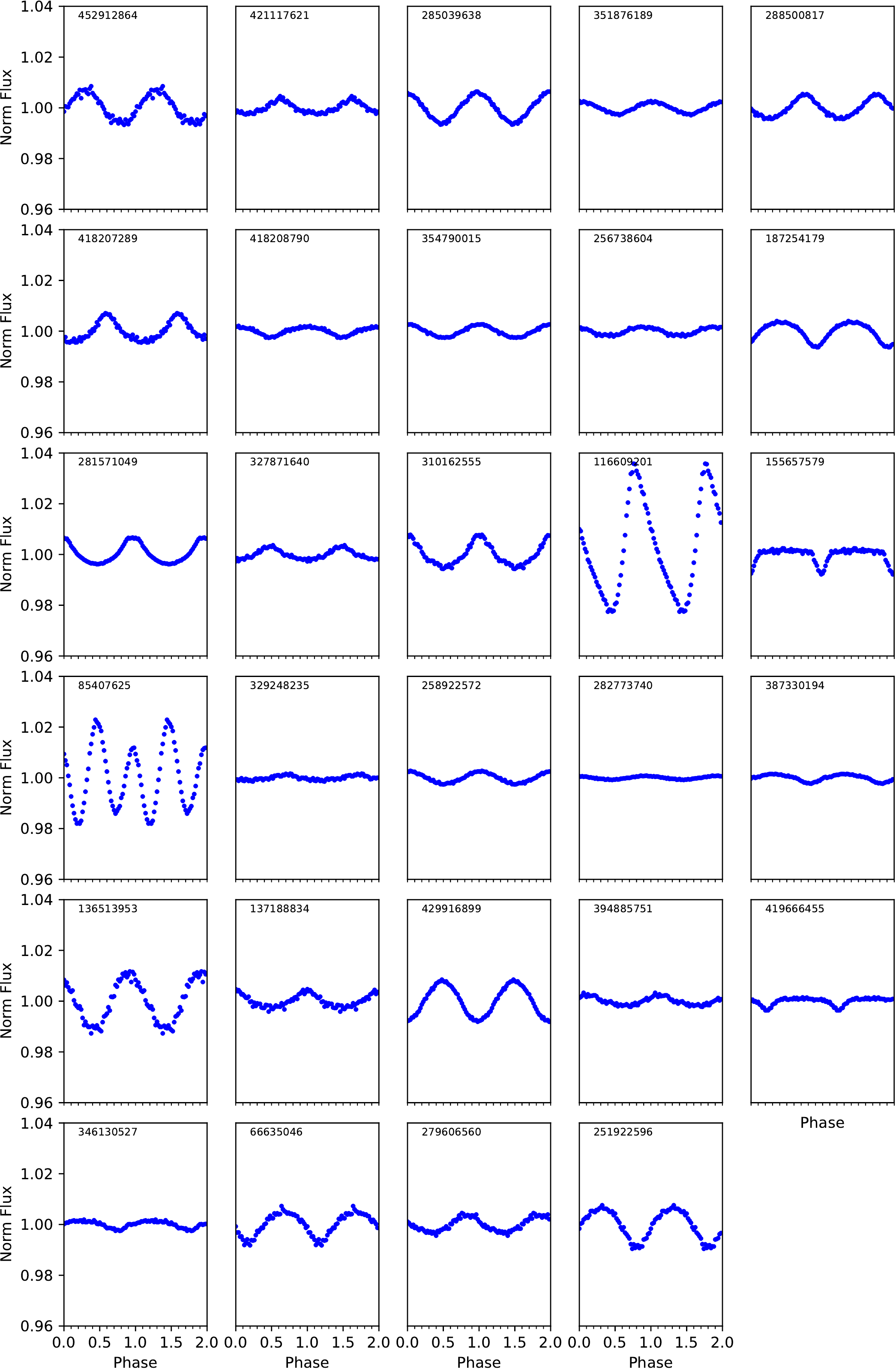}    
\vspace{-2mm}
  \caption{The {\tess} data of the stars in our sample which have been
    folded on the period shown in Table \ref{targets}. To highlight
    the relative amplitude of each star we show the same y-axis scale
    for all sources. Where a star has been observed in more than one
    Sector we show data taken during the first Sector it was
    observed.}
    \label{tessfold}
    \end{center}
\end{figure*}

To gain further insight to the nature of our targets we folded the
data on the period shown in Table \ref{targets}. These folded light
curves are shown in Figure \ref{tessfold} with the semi-amplitude
expressed as fraction indicated in Table \ref{targets}. The vast
majority of our targets show a low amplitude modulation which, at face
value, appear consistent with the presence of starspots which emerge
into and out of view as the star rotates.

There are, however, some stars which have folded light curves which
are clearly not signatures of stars with starspots. TIC 116609201
shows a light curve (and period) which is similar to that of $\delta$
Sct stars or related pulsating stars. There is a relatively bright
star ($G$=12.0) which is 50.2$^{''}$ distant from TIC 116609201. Data
from Gaia DR2 indicates it lies in a position on the HRD which is
consistent with it being a $\delta$ Sct or related pulsator. Given the
size of the {\tess} pixels (21$^{''}$/pixel) some of the light from
this star likely contaminates the light from TIC 116609201. TIC
155657579 and TIC 419666455 appear to be eclipsing binaries with a
period of 4.42 h and 4.73 h respectively (these could be faint
  eclipsing binaries which are spatially nearby the target star). TIC
251922596 may show an eclipse like feature at phase minimum. One
further source, TIC 85407625, shows a more complex light curve, with
two peaks and two minima per cycle. We discuss the nature of this
source in more detail in \S \ref{tic854}.

To search for other targets which may have been affected by light from
other stars we made a systematic search for stars within 1$^{'}$ of
our targets using {\tt tpfplotter} \citep{Aller2020}. (Although the
FWHM of the {\tess} PSF is 1.9 pixels, the number of pixels which are
used to extract the lightcurves in the {\tess} pipeline are typically
3--4). For those stars within our search radius we placed them on the
Gaia HRD \citep{Gaia2018}. We then searched for stars which lay close
to the location where $\delta$ Sct or SX Phe stars lie: these are one
of the few types of variable star which show periodic pulsations on a
period between 0.1--0.2 d. To affect the light curve of our target we
required the nearby star to be at most 1 mag fainter. We find that up
to 1/4 of our targets may have light curves which have been influenced
to a degree by stars which were spatially nearby and which fall in the
Gaia HRD which is consistent with the location of $\delta$ Sct or SX
Phe stars. However, none of these light curves have a shape which is
similar to a classical $\delta$ Sct profile suggesting the effect is
minimum. Further high cadence observations with higher spatial
resolution would be required to identify their location. We note that
although pulsations from low mass stars have been predicted, their
amplitudes are expected to have a fraction of a few 10$^{-6}$
\citep{Rodriguez2019}.

\subsection{Short Period Binaries}

In the previous section we highlighted the need for an examination of
phase folded light curves and also an assessment of the immediate
field for variable stars which could affect the {\tess} photometry of
the target. This led to the identification of two, possibly three,
eclipsing binaries. TIC 155657579 and 419666455 show folded light
curves which are consistent with a low mass - low mass binary. They
have periods which we associate as being the binary orbital period
which are 0.184 and 0.197 d respectively. Such short period low mass
binaries are rare, with only a few being known around the 0.2 d period
\citep[c.f.][]{Zhang2019,Fang2019}. Given we find no evidence for a
variability of the radial velocity of these stars we expect that these
binaries are not related to the low mass star we obtained spectra for
and are likely spatially nearby sources.

\section{Spectroscopic Observations}

We obtained spectroscopy of our targets using the 2.56m NOT on La
Palma for three contiguous nights starting on 27th Oct 2020 using
ALFOSC. We used Grism\# 8, which covers the wavelength range
$\sim$5680-8580~\AA, using a 0.5$^{``}$ slit, giving a spectral
resolution of $R\sim$2000. Exposure times were 300 sec with three
spectra being taken consecutively and later combined to form a single
spectrum after the reductions. For most targets we obtained one set of
three spectra on each of the three nights. These were followed by HeNe
arc lamp exposures. The spectra were bias-subtracted and flatfielded
using the Halogen lamp exposures. The spectra were then extracted
using the {\tt OPTSPECEXTR}
package \footnote{\url{https://physics.ucf.edu/~jh/ast/software/optspecextr-0.3.1}},
that performs optimal extraction of spectra along the lines described
in \citet{Horne1986}. The spectra are shown in Fig \ref{allspec}.

\begin{figure}
  \begin{center}
  \includegraphics[width=0.5\textwidth]{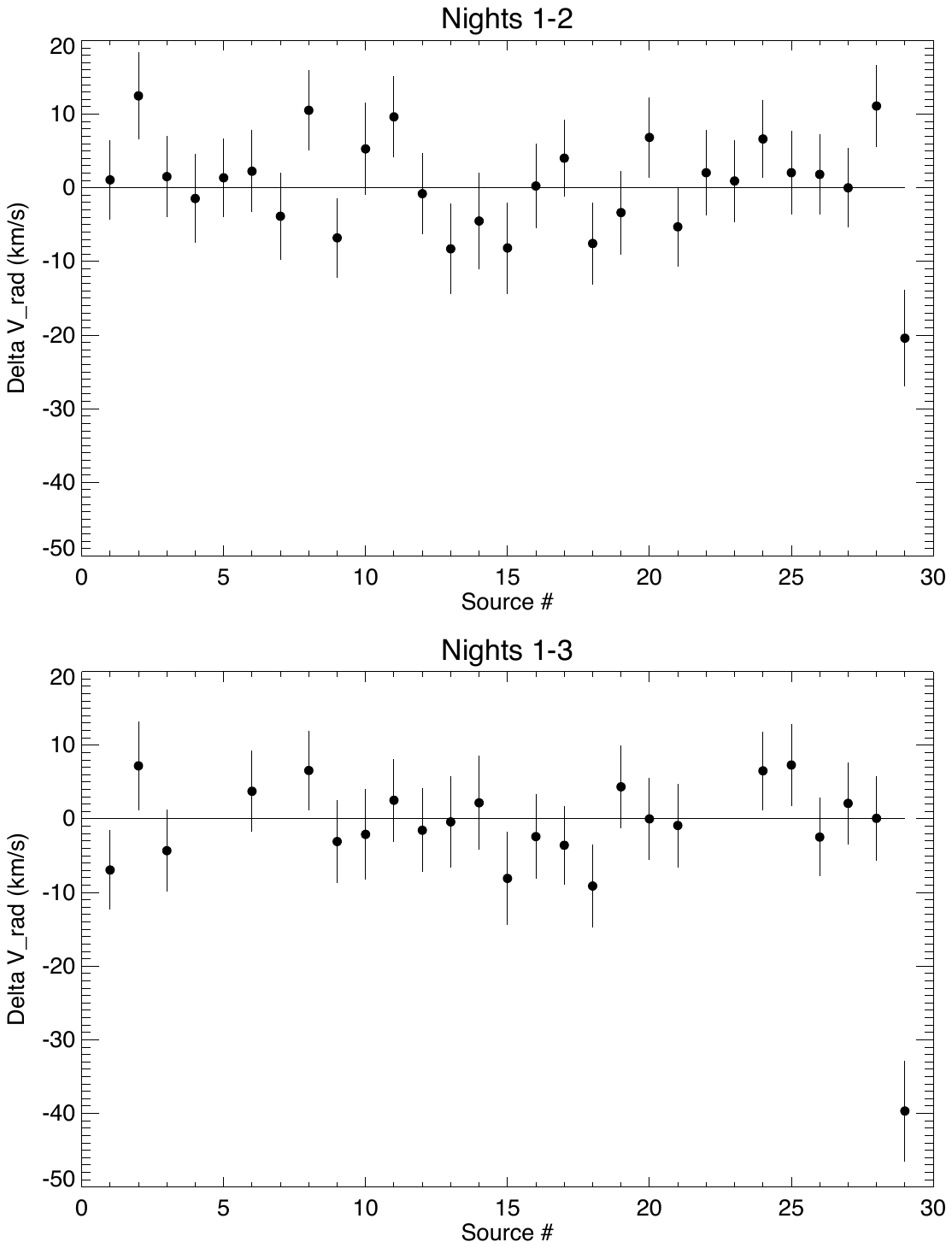}    
\vspace{-2mm}
  \caption{Radial velocity shifts for the 29 targets between nights 1
    and 2 (top) and for the 23 targets that have a third epoch
    spectrum nights 1 and 3 (bottom). The source \#29 (TIC 85407625)
    shows clear changes, whilst the other variations are within the
    noise range.}
    \label{vrads}
    \end{center}
\end{figure}

\begin{figure*}
  \begin{center}
  \includegraphics[width=0.99\textwidth]{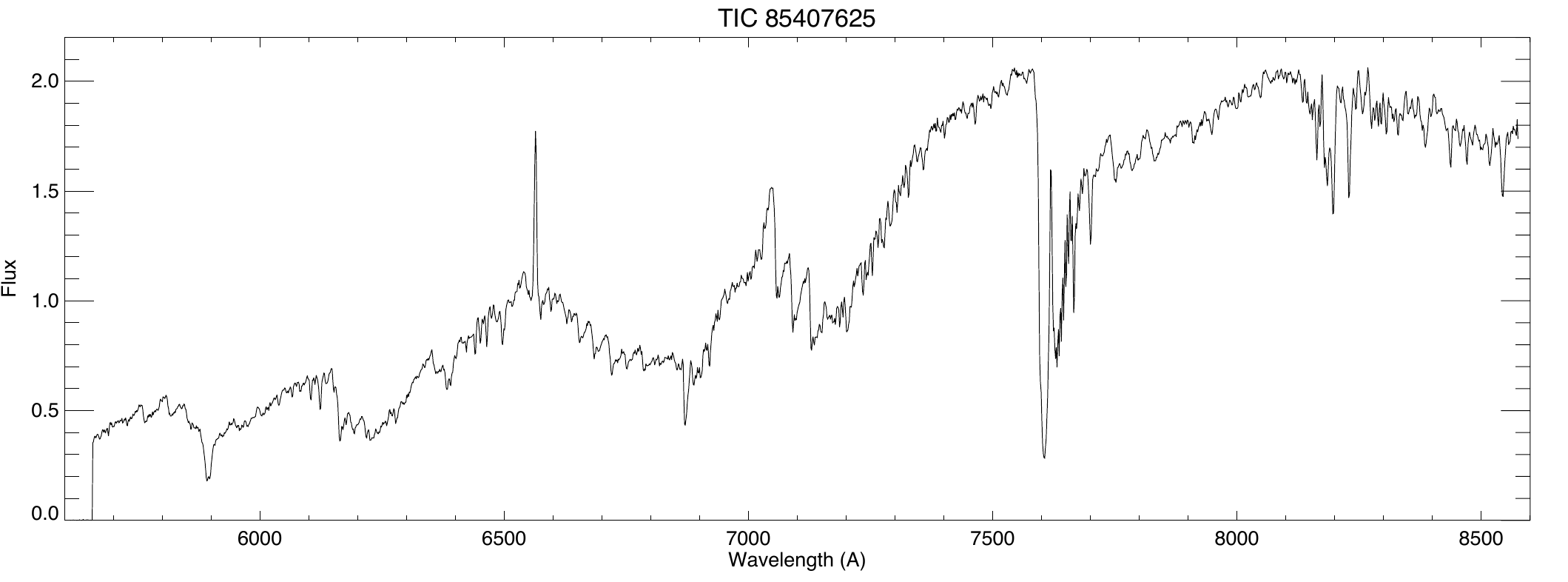}    
  \caption{A spectrum of TIC 85407625 showing clear H$\alpha$ emission
    superimposed on an $\sim$M3V spectrum}
    \label{tic8540spec}
    \end{center}
\end{figure*}

\begin{figure}
  \begin{center}
  \includegraphics[width=0.5\textwidth]{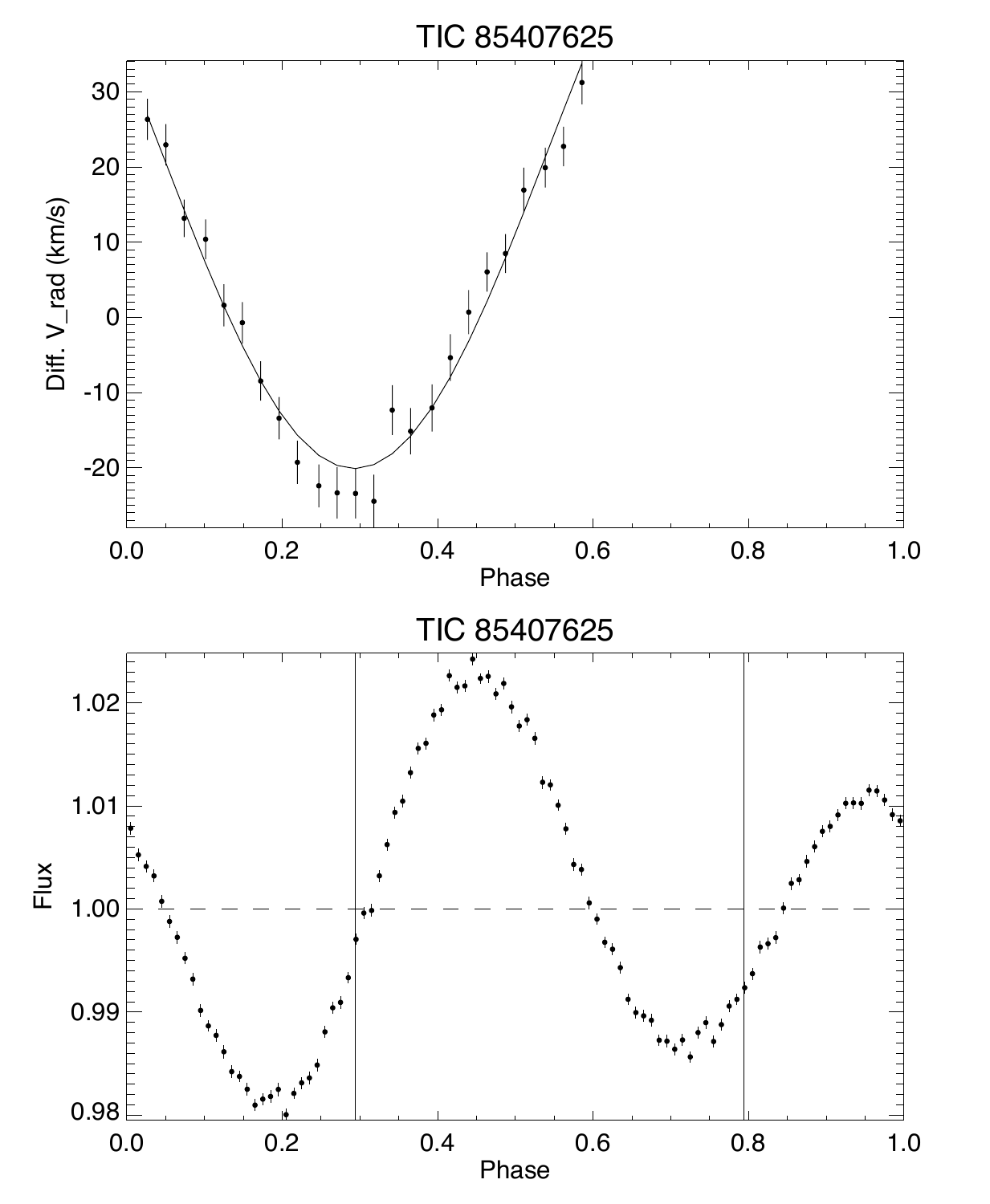}  
\vspace{-2mm}
  \caption{Upper Panel: Radial velocity curve of TIC 85407625,
    together with the best fitting sinusoid which has a period of
    0.1764 d. The radial velocities were determined using a
    cross-correlation of multiple absorption lines. Lower Panel: The
    {\tess} data folded and binned on the same ephemeris (which is
    sufficiently accurate to phase the {\tess} and NOT datasets to
    within $\sim$0.025 (1$\sigma$) phase cycles). The two vertical
    lines mark the phases of maximum blue and (predicted) red shifts.}
    \label{vradcurve}
    \end{center}
\end{figure}

\begin{figure}
  \begin{center}
\includegraphics[width=0.36\textwidth]{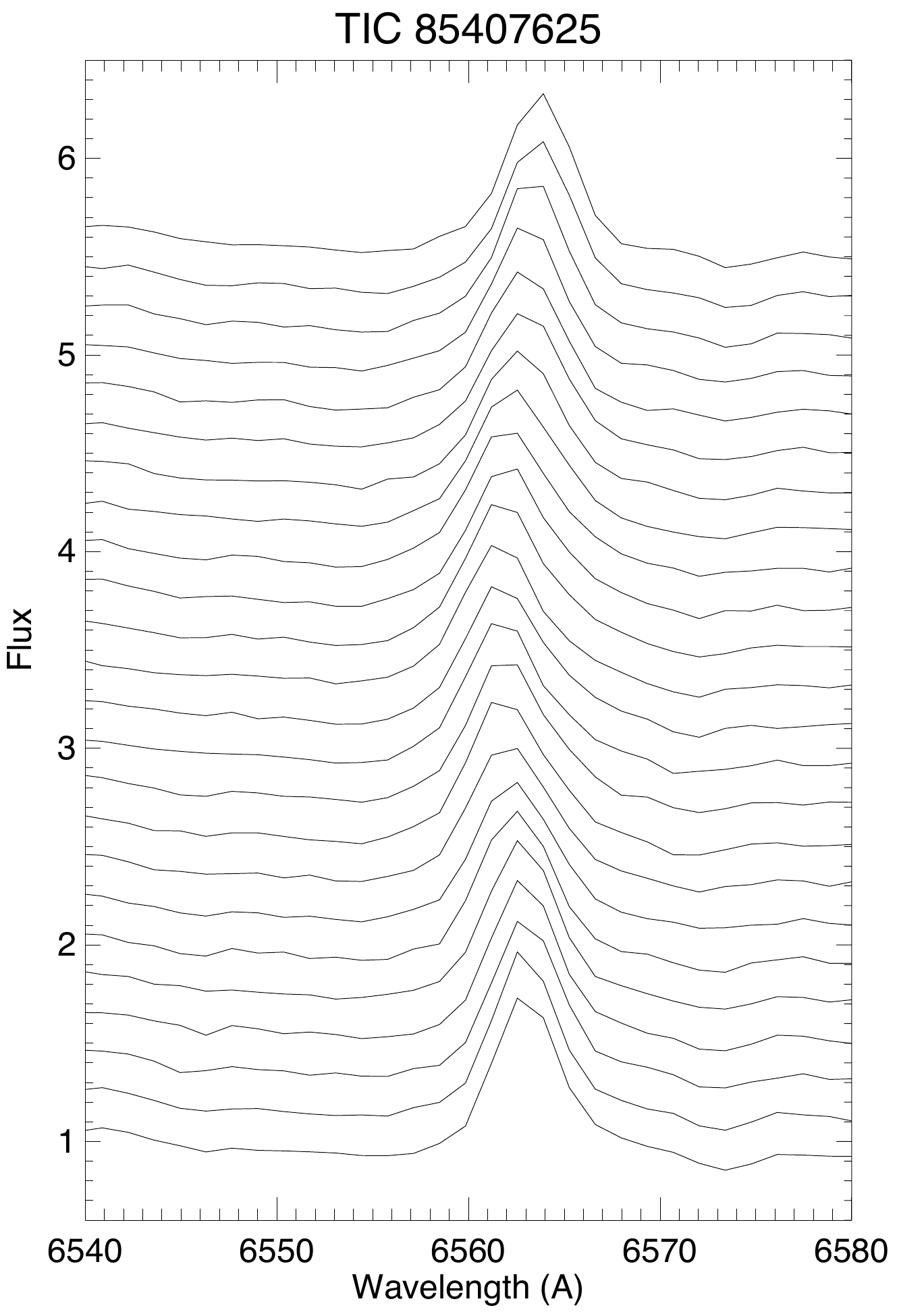}    
  \caption{The spectrum of TIC 85407625 highlighting the change of wavelength of the peak of the H$\alpha$ emission line over time (which runs from top to bottom).}
\label{Halphaspec}
    \end{center}
\end{figure}

\begin{figure}
  \begin{center}
  \includegraphics[width=0.45\textwidth]{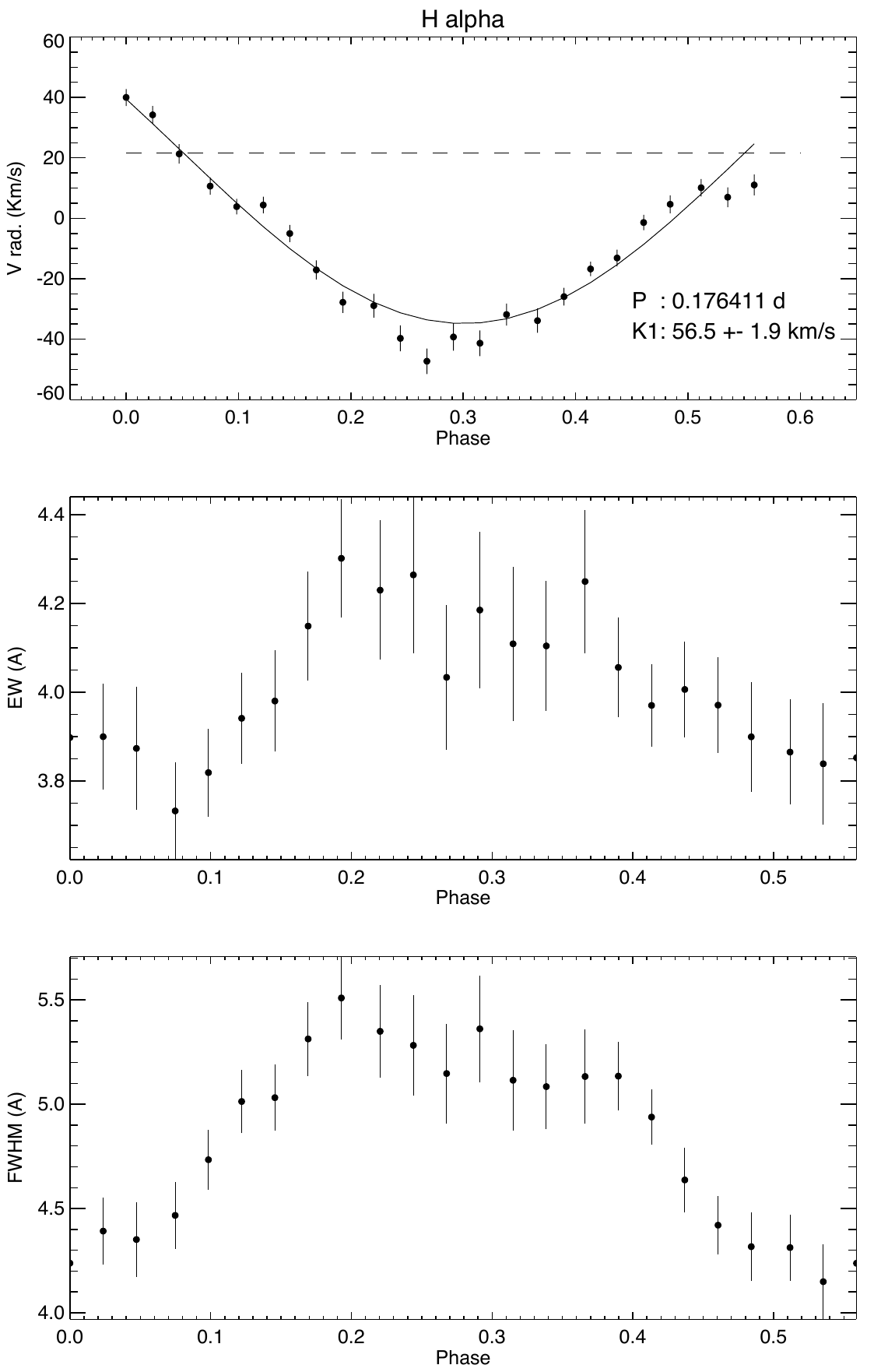}    
  \caption{The radial velocity modulation (top panel); equivalent width (middle panel) and FWHM (lower panel) of the H${\alpha}$ emission line in TIC 85407625.}
\label{Halphavar}
    \end{center}
\end{figure}

\subsection{Searching for Radial Velocity Variations}

In order to search for radial velocity changes in each of the sources,
we cross-correlated the combined spectra of each target from the
second and the third nights against the combined spectrum from the
first night. This was carried out in two specific regions, where M
dwarfs show sharp features in their spectra (i.e. 7000-7160 \AA\ and
8300-8600 \AA). We used an average radial velocity shift from these
two bands. To fully utilise the information content of the spectra, we
interpolated the spectra by a factor of three using splines before
cross-correlating them. In order to correct for any changes in the
instrumental wavelengths more accurately than using the arc lamp
spectra, we cross-correlated two different regions in the spectra that
contain sharp telluric features (i.e. 6840-6890~\AA\ and
7570-7630~\AA). This correction was computed separately for each pair
or source spectra from the nights 1-3. As a result, we obtained
spectra from three nights for 24 sources and from two nights for the
remaining 5 sources. These triplets/pairs of spectra where
cross-correlated in order to search for any evidence for radial
velocity shifts. The typical error for the velocity shifts from
cross-correlation is 6 km s$^{-1}$. This also agrees with the standard
deviation of detected radial velocity shifts for the 28 stars that do
not show any significant velocity shifts. The results are shown in Fig
\ref{vrads}. We detect only one source (TIC 85407625), that shows
clear changes in its radial velocity between different nights.

\subsection{TIC 85407625}
\label{tic854}

In order to further investigate the radial velocity shifts in TIC
85407625, we obtained a time series of spectroscopic observations with
the same instrumental setup as was used for our survey of all 29
targets. The observations took place on 2nd July 2021.  The spectra
were cross-correlated against the first spectrum of the sequence using
the 8300-8600~\AA\ wavelength range (see Fig. \ref{tic8540spec} for
the full spectrum of TIC 85407625). At the time of the observations it
was thought that based on the {\tess} periodogram, the orbital period
of the system was 0.088 d. However, once the phase-folded data
revealed unequal depths for every second maxima and minima, and the
results of a Bayesian analysis of the radial velocity curve were taken
into account, it became clear that the true (orbital) period was
0.1764 d.  We show the resulting radial velocity variation as a
function of orbital period in Fig \ref{vradcurve}. We note that
  for binaries with orbital periods \ltae 4 d it is likely that the
  rotation period of each star in the binary is synchronised or close
  to being synchronised \citep{Lurie2017,Fleming2019}. It is therefore
  likely that the 0.1764 d period is both the binary orbital period
  and the rotation period of the binary components.

To determine the amplitude of the radial velocity variation, we assume
any companion at such short period would have a circular orbit and we
can fix the period at 0.1764 d. The resulting radial velocity curve,
together with a sinusoidal fit, are shown in Fig. \ref{vradcurve}. We
determine a K velocity of 42.6 $\pm$ 1.5 km s$^{-1}$: the $\gamma$
velocity is not calibrated.  This yields a mass function of 58 M$_{\rm
  Jup}$ for the companion star (i.e. a minimum mass for
$i$=90$^{\circ}$). Since a random inclination has a 50\% chance of
being above 60$^{\circ}$, there is a 50\% probability that the
companion mass is $<$90 M$_{\rm Jup}$. There is no sign of a companion
in the spectra, strengthening the case for the brown dwarf
companion. However, a very late type ($\sim$M7+) companion cannot be
ruled out.  We note that the source appears in the catalogue of NIR
spectroscopic survey of nearby M dwarfs \citep{Terrien2015} with a
radial velocity of --3.6$\pm$4.7 km s$^{-1}$ and it is slightly
overluminous for its spectral class.

The spectra of TIC 85407625 also shows strong H$\alpha$ emission (see
Fig. \ref{tic8540spec}), indicative of stellar magnetic activity. The
radial velocities of the H$\alpha$ line (Fig. \ref{Halphaspec}) follow
the motion of the multiple narrow absorption lines used for the radial
velocity measurements, thus confirming the association with the M
dwarf. In addition to the radial velocity modulation (tracking the
movement of the M dwarf), we determined the FWHM and EW of the
H$\alpha$ emission line, which also shows evidence for variation
consistent with the orbital period (Fig. \ref{Halphavar}). This was
estimated by fitting a Gaussian line profile to each of the individual
spectra.  The maximum in both FWHM and EW occurs during the phase of
maximum blue shift. The resulting K velocity from the H$\alpha$ is
somewhat larger (56.0$\pm$1.9 km s$^{-1}$) then the value obtained
from the cross-correlation analysis of the absorption features (42.6
$\pm$ 1.5 km s$^{-1}$). However, as the H$\alpha$ line also shows
changes in its shape and is much wider, we do not consider the
discrepancy between the two values as a serious issue. We do have to
be careful interpreting these results though, since we have
observations only covering 0.6 in orbital phase, even if we can
relatively safely assume an underlying sinusoidal modulation for the
radial velocity.

\section{Discussion}

The underlying aim of this study (and Paper I) is to address the issue
of why some late type stars show evidence of rapid rotation ($<$0.2 d)
but little (or no) flare activity. In particular, our observations
made using the NOT aimed to determine if these stars could be
components of short period binaries.

It is clear from our analysis of the 29 targets outlined in this
paper, that none apart from TIC 85407625 show any
clear evidence for binarity in their radial velocity
measurements. However, given the limited size of our survey population
and the small number of radial velocity measurements per source, we
now set out to place upper limits on the mass of any second binary
component to these stars.

\subsection{Evidence for low mass companions}

\begin{figure}
  \begin{center}
  \includegraphics[width=0.45\textwidth]{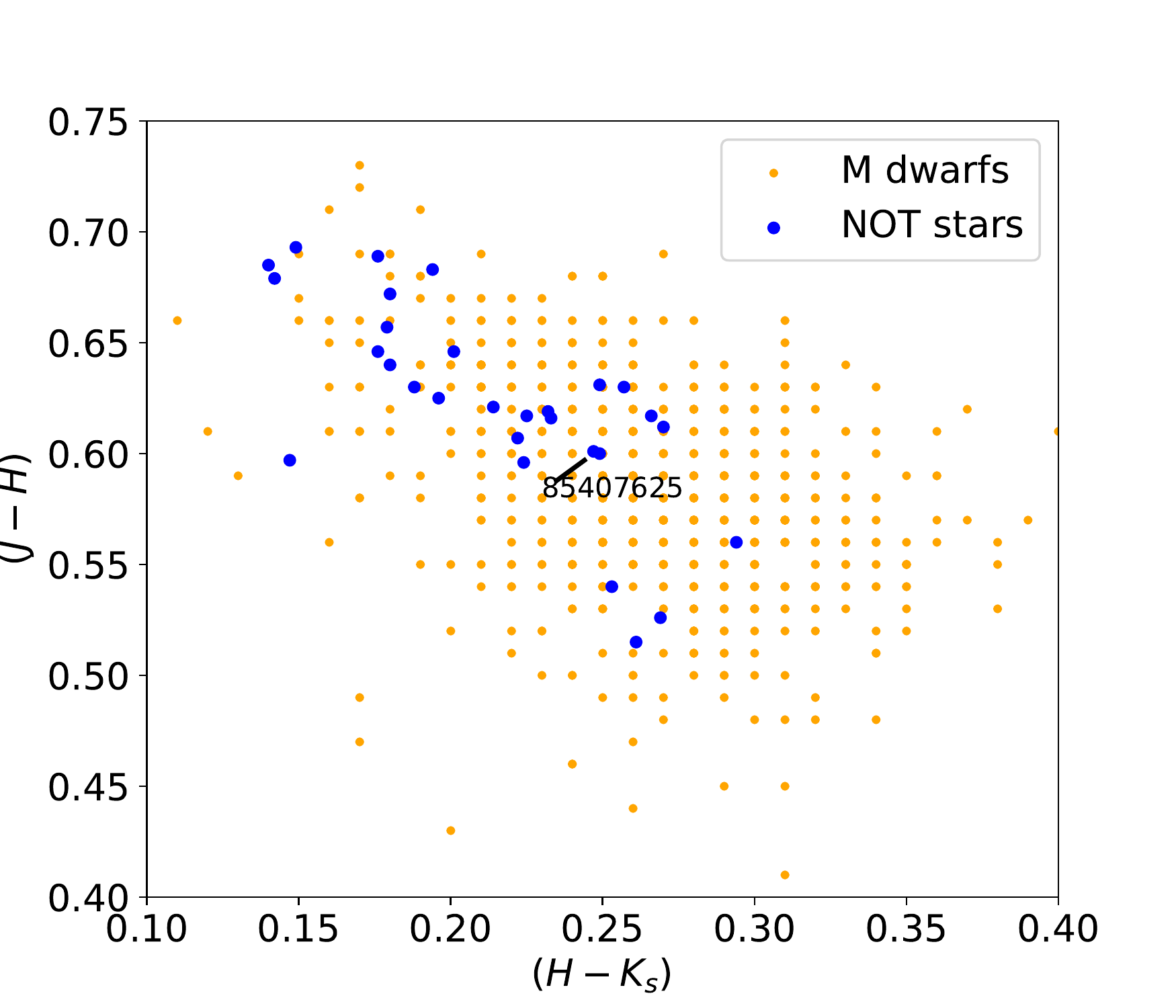}    
\vspace{2mm}
  \caption{The sample of nearby M dwarfs from \citet{Terrien2015} in
    the $(H-K_{s}), (J-H)$ colour-colour plane together with the
    colours of the stars in our NOT sample also obtained using 2MASS
    data. We have noted the position of TIC 85407625.}
    \label{2mass}
    \end{center}
\end{figure}

To explore the possibility that any potential binary companion star to
our targets is visible in the near IR we extracted their 2MASS
$J,H,K_{s}$ magnitudes \citep{Skrutskie2006}. We compare their colours
to the sample of nearby M dwarfs of \citet{Terrien2015}. We show their
colours in Figure \ref{2mass}: the colours of the stars in our NOT
sample are entirely consistent with nearby M dwarfs. Furthermore, if
we compare the absolute $JHK$ magnitude of a M3.5V star (the latest
spectral type of our sample, c.f. Figure \ref{gaiahrd}) with a M8V
star (there is a 50\% probability that the companion mass of TIC
85407625 is below 90 M$_{\rm Jup}$ = 0.084 \Msun), there is a
difference of 3.1, 3.0 and 2.8 mag in their absolute mag. We therefore
do not expect such companions to be detected in $JHK$ colours.

\subsection{Searching for binarity using the Gaia RUWE parameter}
\label{gaiaruwe}

The Gaia data releases \citep{Gaia2018,Gaia2021} incorporate a
  parameter called the Renormalised Unit Weight Error (RUWE) which is
  a measure of how much the photo-center of a star moves over the
  course of the Gaia observations \citep{Lindegren2021a}. Initial
  indications suggest that for stars with RUWE$>$1.4 the star is an
  unresolved binary system \citep{Lindegren2021b}. Gaia EDR3
  \citep{Gaia2021} show that two sources (TIC 137188834 and TIC
  279606560) have RUWE values significantly above 1.4 (6.6 and 11.8
  respectively), suggesting these stars are wide binaries. TIC
  85407625 which shows a radial velocity variation in our NOT data has
  RUWE=1.35, although given the low mass for the secondary star it is
  not clear what its effect on the RUWE value would
  be. \citet{StassunTorres2021} present evidence that even for stars
  which have RUWE=1.0--1.4 may also be unresolved binaries: all but
  one of our targets have RUWE$>$1.0. At this stage there is strong
  evidence that one of our targets is a binary (TIC 85407625), and
  that an additional two stars are possible binaries.

\subsection{Limits on the fraction of binaries}

Although a secondary star to our targets may not be expected to be
seen in near IR colours we now simulate the expected radial velocity
variations which we would have expected to detect given our sampling
rate and an underlying binary model. In our simulation, we exclude the
five sources that only have spectra from two epochs and are therefore
left with 24 sources with spectra from three epochs.

We simulated 10,000 datasets, each of which contained 24 sources,
drawn from a pool of random binaries. These random binaries have
random inclinations; an orbital period between 0.1-0.2 d; M dwarf
masses between 0.3-0.6M$_\odot$ (to match the spectral classes of our
sample) and secondary masses between 10-80 M$_{Jup}$. The values are
uniformly distributed (apart from the inclination). The three
observation epochs are taken at random orbital phases. As a result we
find that, on the average, we should observe 2.3 sources that show one
radial velocity difference $>$20 km s$^{-1}$ and another with $>$40 km
s$^{-1}$: the distribution is Poissonian. We find that in 30\% of
cases we would observe 0 or 1 sources, with radial velocity shift
detections at the level which we observed in TIC 85407615. However,
the scatter in the radial velocity shifts of the remaining sources is
so small, that it cannot originate from the aforementioned
distribution.

The two sample Kolmogorov-Smirnov test yields a $2.0\times10^{-8}$
probability for our sample to originate from the simulated binary
distribution described above. It is therefore not feasible that the
sources in our sample could originate from the underlying distribution
of M dwarf-brown dwarf binaries (and even less feasible that they
could have late type M companions) and most of them are likely to be
single M dwarfs.

\subsection{Amplitude of modulation}

In Table \ref{targets} we note the semi-amplitude of the modulation
seen in {\tess} light curves of our targets. If we omit TIC 116609201
(contamination from a $\delta$ Sct star); TIC 155657579 and TIC
419666455 (contamination from an eclipsing binary) and TIC 85407625
(the binary system noted in \S \ref{tic854}) we find they have a mean
fractional semi-amplitude of 0.00752. In \citet{Ramsay2020} we
explored the flare rate of low mass stars in {\tess} Cycle 1
data. There were only 6 stars which had rotational periods $<$0.2 d
and were classed as flare active (at a rate $>$0.044 flares/day). The
mean fractional semi-amplitude of these stars is 0.0375. Using the
two-sample Anderson-Darling test we find the two samples differ at the
3$\sigma$ level: we therefore find marginal evidence that the flare
in-active UFRs reported here have a smaller amplitude of modulation
than flare active UFRs.  Our sample of stars with NOT spectra show no
optical flares; show no evidence of H$\alpha$ in emission and appear
to also show some evidence of having a comparatively lower amplitude
of modulation in their {\tess} light curves.  It is possible that a
signature of their rotation period was only possible because of the
sensitivity of {\tess} and that they have starspots covering a
relatively small fraction of the stars photosphere. 
  Alternatively, they could have a large number of small spots which
  are widely distributed over the star \citep{JacksonJeffries2013},
  and see \citet{Luger2021} for a recent paper outlining the
  difficulties of predicting the distribution of spots from single
  band light curves.

\subsection{Super-saturation?}

The lack of any flares in these systems is puzzling, as such fast
rotation should generate magnetic activity due to the dynamo process,
which should manifest itself also in the form of flares. However, as
noted in our Paper I \citep{Doyle2022}, this might be explained by the
effect thought to be behind the supersaturation of X-ray emission in
the fast rotating M dwarfs i.e. the fast enough rotation opens up the
magnetic loops in the corona due to centrifugal force
\citep{Jeffries2011}, which disables the storage of magnetic energy in
the coronal loops. As a consequence, the magnetic reconnection is
inhibited and no flares are detected, even if star spots are observed
in the photosphere.  In early M dwarfs, the centrifugal force is a
  factor of four greater for a star with a rotation period of 0.15 d
  compared to 0.3 d. However, the centrifugal force is also a factor
  of five times stronger for a star of spectral type M0 compared to
  M4. We are searching for more low mass UFRs in {\tess} data to
  attempt to disentangle these effects.

\section{Conclusions}

We have shown that in studies which explore the activity
levels of low mass stars, it is essential to examine their phase folded
light curves. A small number show light curves where the variability
is clearly not due to starspots. Further examination shows that around
1/4 of our targets may have light curves which could have been
effected to a small degree by stellar pulsators, such as $\delta$ Sct
stars, which are spatially nearby our target. However, the majority of
our targets appear to be consistent with being low mass stars in which
the variability is due to starspots.

We have set out to address the question of why a significant number of
low mass stars which have a periodic modulation in their {\tess} light
curve $<$0.2 d do not show clear evidence for flare activity.  We have
presented the results of a radial velocity survey of 29 stars which
show a periodic modulation in the {\tess} light curve. One star
  shows a clear radial velocity variation with another two stars
  showing high RUWE values (\S \ref{gaiaruwe}) which suggests that
  they too maybe binary systems.

In Paper I, where we reported that half the sample of ten stars showed
evidence for a line of sight magnetic field strength of
$\sim$1--2~kG, we suggested that the lack of flares in UFRs which also
show evidence for a magnetic field, could be related to
super-saturation which inhibits magnetic reconnection and hence the
production of flares. In \citet{Ramsay2020} we suggested that stars
whose activity is saturated could produce flares through micro-flares
in the $U$ band. Observations of flare inactive UFRs using high
cadence instruments on mid-sized telescope which have $U$-band
sensitivity are encouraged. In this paper we note that UFRs which are
not flare active appear to show some evidence for a lower modulation
amplitude to flare active UFRs. Further observations of low-mass stars
using {\tess} data (especially the 20 sec cadence mode introduced in
Cycle 3) should help address this question. Also, medium resolution
spectroscopic observations could reveal the presence of Doppler
broadened lines which would be expected from rapidly rotating
stars. Finally, we report the discovery of a possible M3 dwarf - brown
dwarf binary TIC85407625 with the mass function of 58 M$_{\rm Jup}$
and an orbital period of 0.176 d. Close M dwarf - brown dwarf binaries
are extremely rare. The companion mass has a 90 M$_{\rm Jup}$ upper
limit with 50\% probability and there is no sign of a companion in the
spectra ruling out a binary M dwarf, although a very late type
($\sim$M7+) companion cannot be ruled out.

\section*{Acknowledgements}

The data presented here were obtained, in part, with ALFOSC, which is
provided by the Instituto de Astrofisica de Andalucia (IAA) under a
joint agreement with the University of Copenhagen and NOT. This paper
includes data collected by the {\tess} mission. Funding for the
{\tess} mission is provided by the NASA Explorer Program.  The Gaia
archive website is \url{https://archives.esac.esa.int/gaia}. Armagh
Observatory and Planetarium is core funded by the Northern Ireland
Executive through the Dept. for Communities. LD acknowledges funding
from a UKRI Future Leader Fellowship, grant number MR/S035214/1. JGD
would like to thank the Leverhulme Trust for a Emeritus
Fellowship. We thank the anonymous referee for a helpful report.

\section*{Data Availability}

{\tess} data are available from the NASA MAST portal. NOT spectra can
be obtained on request from the authors.

\vspace{4mm}

\bibliographystyle{mnras}

\appendix 

\section{Figures}

\begin{figure*}
  \begin{center}
  \includegraphics[width=0.99\textwidth]{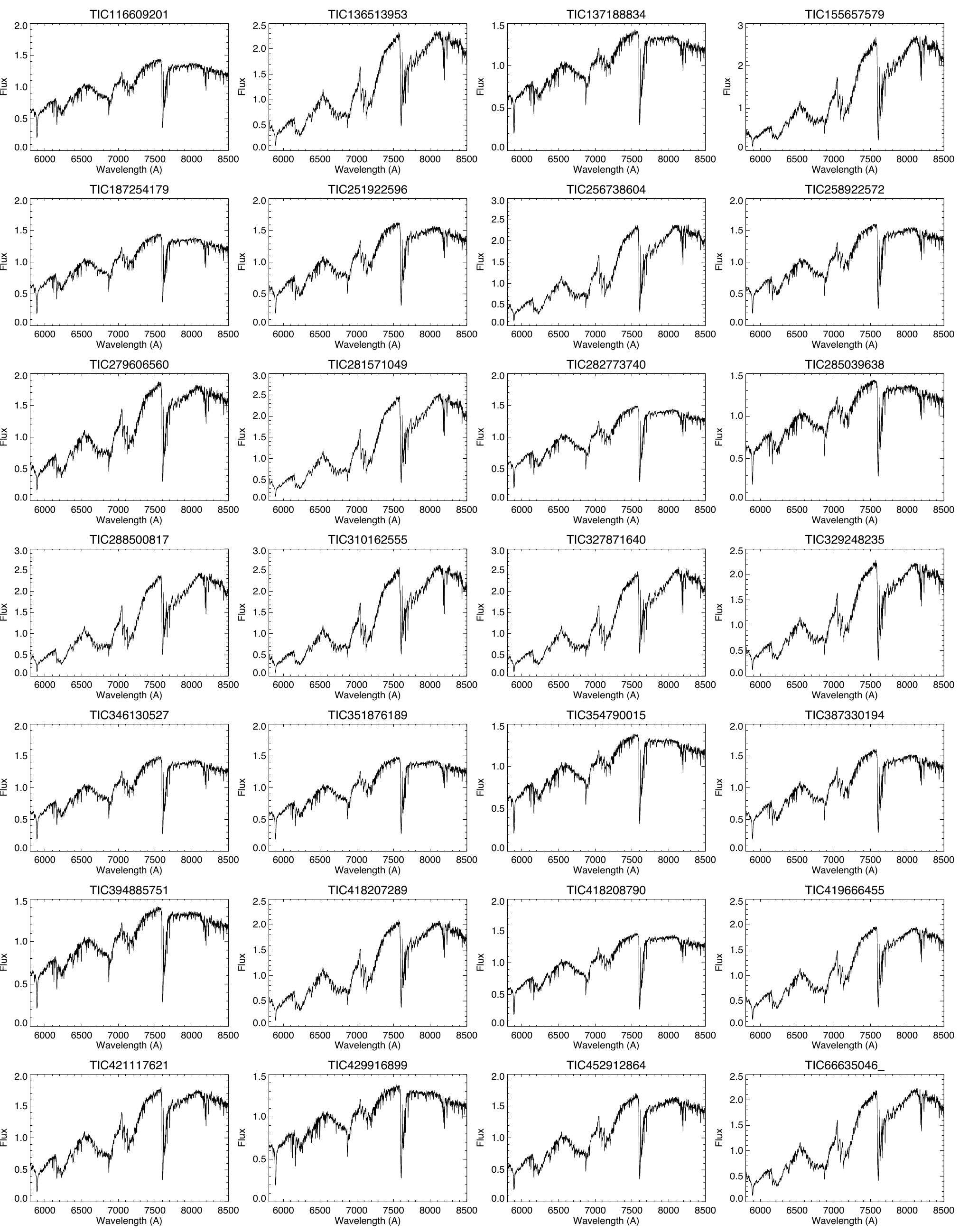}    
\vspace{-2mm}
  \caption{The 28 spectra of our sample (excluding TIC 85407625, shown separately). }
    \label{allspec}
    \end{center}
\end{figure*}

In Figure \ref{allspec} we show the first optical spectra of each of our target stars, with the exception of TIC 85407625 which is shown in Figure \ref{tic8540spec}.

% Don't change these lines
\bsp	% typesetting comment
\label{lastpage}

\end{document}